\documentclass[usegraphicx,usenatbib]{mn2e}
\usepackage{times,amssymb,amsbsy}
\usepackage[fleqn]{amsmath}

\newcommand{\D}[2]{\frac{\partial #2}{\partial #1}}
\newcommand{\DD}[2]{\frac{\partial^2 #2}{\partial #1^2}}
\newcommand{\deriv}[2]{\frac{{\rm d} #2}{{\rm d} #1}}

\newcommand\bb[1]{\mbox{\boldmath{$#1$}}}
\newcommand\grad{\bb{\nabla}}
\newcommand\bcdot{\bb{\cdot}}
\newcommand\btimes{\bb{\times}}
\newcommand{\mc}[1]{\mathcal{#1}}

\newcommand{\msb}[1]{\bb{\mathsf{#1}}}

\newcommand{\imag}{{\rm i}}
\newcommand{\eb}{\hat{\bb{b}}}
\newcommand{\ex}{\hat{\bb{x}}}
\newcommand{\ey}{\hat{\bb{y}}}
\newcommand{\ez}{\hat{\bb{z}}}
\newcommand{\vasq}{v^2_{\rm A}}
\newcommand{\vthsq}{v^2_{\rm th}}
\newcommand{\mfp}{\lambda_{\rm mfp}}
\newcommand{\BB}{\bb{B}}

\title[The HBI in a quasi-global model of the ICM]
{The HBI in a quasi-global model of the intracluster medium}

\author[Latter \& Kunz]
{Henrik N. Latter$^{1}$\thanks{E-mail: hl278@cam.ac.uk}, Matthew W.
Kunz$^{2,3}$\thanks{E-mail: kunz@astro.princeton.edu; Einstein Postdoctoral
Fellow}\\
$^{1}$ Department of Applied Mathematics and Theoretical Physics, University of
Cambridge, CMS, Wilberforce Road, Cambridge, CB3 0WA, U. K.\\
$^{2}$ Department of Astrophysical Sciences, 4 Ivy Lane, Peyton Hall, Princeton
University, Princeton, NJ 08544, U. S. A.\\
$^{3}$  Previous Address: Rudolf Peierls Centre for Theoretical Physics,
University of Oxford, 1 Keble Road, Oxford, OX1 3NP, U. K.
}
\date{}
\pagerange{\pageref{firstpage}--\pageref{lastpage}} \pubyear{2011}
\def\LaTeX{L\kern-.36em\raise.3ex\hbox{a}\kern-.15em
    T\kern-.1667em\lower.7ex\hbox{E}\kern-.125emX}
\begin{document}
\label{firstpage} \maketitle

\begin{abstract}
In this paper we investigate how convective instabilities 
influence heat conduction in the intracluster 
medium (ICM) of cool-core galaxy clusters. The ICM is a high-beta, weakly collisional 
plasma in which the transport of momentum and heat is aligned with the magnetic field. 
The anisotropy of heat conduction, in particular, 
gives rise to instabilities that can access energy stored in a 
temperature gradient of either sign. We focus on 
the heat-flux buoyancy-driven instability (HBI), 
which feeds on the outwardly increasing temperature profile of cluster cool cores. Our aim is to elucidate 
how the global structure of a cluster impacts on the growth and morphology of the linear HBI modes when 
in the presence of Braginskii viscosity, and ultimately on the ability of the HBI to thermally insulate cores. 
We employ an idealised quasi-global model, the plane-parallel atmosphere, which captures 
the essential physics -- e.g. the global radial profile of the cluster -- while letting the problem remain analytically tractable. 
Our main result is that the dominant HBI modes are localised to the innermost ($\lesssim$$20\%$) regions of 
cool cores. It is then probable that, in the nonlinear regime, appreciable field-line insulation will be similarly 
localised. Thus, while radio-mode feedback appears necessary in the central few tens of kpc, heat conduction 
may be capable of offsetting radiative losses throughout most of a cool core over a significant fraction of the Hubble time. 
Finally, our linear solutions provide a convenient numerical test for the nonlinear codes that 
simulate the saturation of such convective instabilities in the presence of anisotropic transport.
\end{abstract}

\begin{keywords}
conduction -- instabilities -- magnetic fields -- MHD -- plasmas -- galaxies:
clusters: intracluster medium.
\end{keywords}

\section{Introduction}

Whereas their copious X-ray emission suggests a central cooling time much less 
than the Hubble time, galaxy cluster cores do not exhibit cooling flows 
commensurate with their radiative losses. Absent are the mass deposition rates 
($\dot{M} \sim 10^2$--$10^3~{\rm M}_\odot~{\rm yr}^{-1}$) and copious 
iron line emission that would accompany such flows. Instead, the central 
temperatures of cool core clusters are 
typically only $\sim$$1/3$ of the bulk cluster temperatures, while 
spectroscopically determined mass deposition rates are 
$\lesssim$$0.1\dot{M}$  \citep[for a review, see][]{pf06}. Understanding how these 
conditions are maintained in the face of rapid radiative cooling poses a challenge for 
theorists, who have consequently invoked a plethora of diverse physics to 
resolve the problem. These include: active galactic nucleus feedback, 
conductive heat transport, and convective 
turbulence, whether it be driven by cosmic rays, mergers, or 
magnetohydrodynamic (MHD) instabilities.

This paper concerns the contribution of conductive heat transport to the 
solution of the cooling flow problem. However, we do not take the usual energetics 
standpoint (i.e. does conduction provide an ample source of heat to offset radiative 
losses?) but rather focus on how conduction generates, and is subsequently modified 
by, plasma instabilities and the MHD turbulence they instigate. Because the intracluster 
medium (ICM) is magnetized and weakly collisional it exhibits surprising stability properties 
\citep{balbus00,balbus01,sckhs05,scrr10}. In particular, the outwardly increasing temperature 
profile of a cluster core gives rise to a heat-flux buoyancy-driven instability 
\citep[HBI;][]{quataert08} that can drive disordered flows. Numerical simulations show that 
these flows lead to near-complete field-line insulation of the core, and heat conduction 
is greatly impeded \citep{pq08,brbp09,pqs09}. HBI motions not only seem incapable of 
maintaining the observed temperature profile (as suggested by \citealt{br08}), 
but actually exacerbate the cooling flow problem. As a result, theorists have 
appealed to turbulent stirring by other means to reopen the field lines 
and reinstate the conductive heat-flux \citep{pqs10,ro10,mpsq11}.

This would bring the issue to a close if it were not for the fact that a magnetized, 
weakly collisional plasma can exhibit appreciable pressure anisotropy 
 -- an effect neglected in most previous work. In fact, HBI modes inevitably generate 
 such anisotropies, manifested as viscous stresses, 
and these adversely affect the instability mechanism. 
In particular, the wavelengths of the fastest growing modes 
$\lambda_{\rm{HBI}}$ increase from very small scales to those of order the 
thermal-pressure scaleheight of the cluster $H$ \citep{kunz11}.
So, even though the HBI mechanism can be understood with local physics,
 the instability manifests primarily on \emph{global} scales 
and may be sensitive to the 
large-scale structure of the cluster, especially via the kinematic viscosity's strong 
dependence on temperature and density. 
For this reason, \citet{kunz11} conjectured, on account of a WKBJ analysis,
 that the fastest growing HBI modes would favour only 
the innermost regions of the cluster core,
as it is here that viscous damping is minimised.
If the nonlinear saturation of the HBI is similarly confined, then
the associated field-line insulation of the core may be significantly attenuated.

In this paper, we investigate how pressure anisotropy and the global 
structure of the ICM influence the HBI 
by generalising the linear calculation of \citet{kunz11}. We calculate the HBI 
modes in a simple quasi-global model of the cluster: a plane-parallel 
atmosphere. The model captures the radial density and temperature structure of 
the ICM, but neglects its angular structure. 
Though idealized, this model permits us to test in a transparent way the new 
physics at work, and how the character of the HBI changes in a global setting. 
In addition, it provides a numerical testbed for nonlinear codes that 
include anisotropic conduction and viscosity. Our main result is that the 
dominant HBI modes are typically localised to the innermost ($\lesssim$$20\%$) 
regions of cool-core clusters. It is then likely that, in the nonlinear regime, 
appreciable field-line insulation will similarly localise and become inefficient.
These theoretical results have been corroborated by the Braginskii-MHD simulations 
presented in \citet{kbrs12}, which use domains sufficiently 
large to capture this physics.

The paper is organized as follows. In Section \ref{sec:formulation} we present 
the model and its governing equations. In Section \ref{sec:equilibrium} we 
derive quasi-global equilibrium solutions to these equations, which are then 
subjected to linear perturbations in Section \ref{sec:linearshit}. There, the 
HBI eigenmodes are calculated numerically, presented, and discussed. We conclude in 
Section \ref{sec:discussion} with a speculative discussion of the implications 
our results have for the long-term thermal and dynamic stability of cool-core 
galaxy clusters.

\section{Formulation of the Problem}\label{sec:formulation}

\subsection{Plasma dynamics}\label{sec:plasmas}

The macroscopic scales of the ICM (e.g. the 
thermal-pressure scale height $H$) are much greater than the particle mean free 
path $\mfp$ and thus a 
(collisional) MHD description of the ICM plasma remains valid. On the other hand,
$\mfp$ is 
much larger 
than the ion gyroradius, a physical regime that ensures the predominance of 
cyclotronic motions. As a consequence, the transport 
properties of the plasma deviate significantly from what we might expect 
from classical MHD. 
For instance, the conduction of heat is strongly anisotropic with respect to 
the local magnetic field direction, a property that fundamentally alters the 
ICM's convective stability \citep{balbus00}. 
Temperature gradients, rather than entropy gradients, become the discriminating 
quantities that determine stability, regardless of whether temperature 
increases \citep{balbus00,balbus01} or decreases \citep{quataert08} in the 
direction of gravity. In addition, the fluid is subject to pressure 
anisotropies, i.e. the gas pressure perpendicular $p_\perp$ and parallel 
$p_{\|}$ to the local magnetic field may not be equal. 
Pressure anisotropies arise from the conservation of the first and second 
adiabatic invariants of each particle. These ensure that any change in magnetic 
field strength $B$ and/or density $\rho$ must be accompanied by corresponding 
changes in $p_\perp$ and $p_{\|}$ \citep{cgl56}.

The pressure anisotropy impinges noticeably 
on the buoyancy instabilities that afflict stratified plasmas, such as the 
magnetothermal instability (MTI) and the HBI. Because the anisotropy
impedes the convergence and divergence of magnetic field lines, 
the MTI mechanism is strengthened, while the HBI is suppressed. 
In particular, pressure anisotropy shifts the fastest-growing HBI modes to much 
longer wavelengths than the very short scales it was thought to favour. 
In fact, \citet{kunz11} finds that the fastest growth occurs for a
wavelength $\sim 10 ( \mfp H )^{1/2}$, which in the weakly collisional ICM is 
approximately global, $\sim 0.2H$--$H$. These longer modes operate on a 
timescale slower than conduction but faster than viscous diffusion;
thus they can effectively access free energy  
via the heat-flux, while minimising viscous damping.

Pressure anisotropies also give rise to a host of `microinstabilities', which 
include the firehose, mirror, and gyrothermal instabilities \citep{sckhs05,
scrr10}. These generate turbulent fluctuations on `nano-scales' of 
$\sim$$10~{\rm npc}$ and $\sim$$10~{\rm hr}$, which likely set the pressure 
anisotropy and heat-fluxes on the larger scales that the HBI inhabits 
\citep{scrr10}. Ideally, these fluctuations would be `smoothed out' by a mean 
field theory or else accounted for in some fashion. In this work we assume from 
the outset that our plasma is Maxwellian, and so these instabilities do not appear in our 
linear theory (though they may during the nonlinear phase of the evolution).

\subsection{Governing equations}\label{sec:equations}

The model we use is that of Braginskii-MHD, in which the equations of classical 
MHD are employed but with special prescriptions (closures) for the transport of 
momentum and heat \citep{braginskii65}. 
The evolutionary equations governing the mass density $\rho$, velocity 
$\bb{v}$, dimensionless entropy  $S=\ln (\rho^{-5/3} p)$, and magnetic field 
$\BB$ are, respectively,
\begin{equation}
\frac{D \rho}{D t}  = - \rho \grad \bcdot \bb{v} , \label{eqn:continuity}
\end{equation}
\begin{equation}
\frac{D \bb{v}}{ Dt}  =  \bb{g} - \frac{1}{\rho} \grad \left( \msb{P} +
\msb{I}\,\frac{B^2}{8\pi} -\frac{\BB\BB}{4\pi} \right),
 \label{eqn:momentum}
\end{equation}
\begin{equation}
\frac{3}{2}\,p\,\frac{D S}{Dt}  = \Gamma - \Lambda - \grad \bcdot \bb{q} ,
\label{eqn:pressure}
\end{equation}
\begin{equation}
\frac{D\BB}{Dt}  = \BB \bcdot \grad \bb{v} - \BB \, \grad \bcdot \bb{v},
\label{eqn:induction}
\end{equation}
where
\begin{equation}
\frac{D}{Dt}\equiv \D{t}{} + \bb{v} \bcdot \grad
\end{equation}
is the convective (Lagrangian) derivative, $\bb{g}$ is the gravitational 
acceleration, and $\msb{P}$ is the (thermal) pressure tensor with the isotropic 
pressure defined through
$p=(1/3)\text{Tr}\,\msb{P}$. The 
viscous heating rate is
$\Gamma$, the radiative cooling rate is $\Lambda$, and $\bb{q}$ is 
the conductive heat-flux. 
In addition, the magnetic field must satisfy $\grad\bcdot\BB=0$. Throughout we 
assume that the gas satisfies the ideal gas law $p = \rho \vthsq$, where 
$\vthsq = 2 k_{\rm B} T / m_{\rm p}$ is the thermal speed of the ions 
(with $k_{\rm B}$ denoting the Boltzmann constant and $m_{\rm p}$ proton mass). 
The adiabatic index $\gamma$ has been set to $5/3$. We consider a hydrogenic 
plasma with equal ion and electron number densities, $n_{\rm i} = n_{\rm e}$, 
and temperatures, $T_{\rm i} = T_{\rm e} = T$.

\subsubsection{Pressure tensor}

The ICM plasma distribution function is gyrotropic. 
As a consequence, the pressure tensor reduces to:
\begin{equation}
\msb{P} = p_\perp(\msb{I}- \eb \eb) + p_\|\eb \eb,
\end{equation}
where $\eb= \BB/B$, and $p_\perp$ and $p_\|$ are the pressures perpendicular 
and parallel to the local magnetic field. The total gas pressure is
\begin{equation}
p = \frac{2}{3} p_\perp + \frac{1}{3} p_{\|}.
\end{equation}
When the ion--ion collision time $\tau_{\rm ii}$ is much shorter than the 
characteristic timescales associated with the macroscopic fields, the pressure 
anisotropy may be computed from
\begin{equation}\label{eqn:anisotropy}
p_\perp - p_{\|} = 0.960 \, p_{\rm i} \tau_{\rm ii} \deriv{t}{} \ln
\frac{B^3}{\rho^{2}}.
\end{equation}
Here we have ignored the contribution of the electrons, which is a factor 
$\sim$$( m_{\rm e} / m_{\rm i} )^{1/2}$ smaller than that of the ions \citep{cs04}.
By using equations (\ref{eqn:continuity}) and (\ref{eqn:induction}) to replace 
the time derivatives of density and magnetic field strength with velocity 
gradients, equation (\ref{eqn:anisotropy}) may be written as 
\begin{equation}
p_\perp - p_{\|} = 3\rho \nu \left( \eb\eb - \frac{1}{3} \msb{I} \right)
\bb{:}\grad\bb{v} ,
\end{equation}
where we have introduced the (kinematic) viscosity coefficient 
\begin{align}
\nu & \equiv  0.960 \times \frac{1}{2} v_{\rm th}^2 \tau_{\rm ii}, \\
       & \simeq  0.031 \left(\frac{n_{\rm i}}{0.01~{\rm cm}^{-3}}\right)^{-1}
\left(\frac{k_{\rm B} T}{2~{\rm keV}} \right)^{5/2}~{\rm kpc}^2~{\rm Myr}^{-1}
. \nonumber
\end{align}
This pressure anisotropy is the physical effect behind what is known as 
\citet{braginskii65} viscosity -- the restriction of the viscous damping 
 to the motions and gradients parallel to the magnetic field.

Other than impeding motions in the momentum equation \eqref{eqn:momentum}, the 
Braginskii stress appears in the entropy equation (\ref{eqn:pressure}) through 
the viscous heating term, $\Gamma$. Using the closure \eqref{eqn:anisotropy}, 
this heating term is $\propto \nu_{\rm ii} ( p_\perp - p_{\|} )^2$ and is therefore higher 
order in the linear analysis we perform.

\subsubsection{Heat conduction}

When the particle gyroradius is much smaller than the collisional mean free 
path, heat is restricted to flow along magnetic lines of force 
\citep[e.g.][]{braginskii65}. The heat-flux can then be written as
\begin{equation}
\bb{q} = - \rho \kappa \, \eb \eb \bcdot \grad \vthsq.
\end{equation}
The parallel thermal diffusivity $\kappa$ is dominated by the contribution from 
the electrons,
\begin{eqnarray}\label{eqn:kappa}
\kappa & \equiv & 1.581 \times \frac{1}{2} v^2_{\rm th,e}  \tau_{\rm ee} \\*
& \simeq & 1.67 \left( \frac{n_{\rm i}}{0.01~{\rm cm}^{-3}} \right)^{-1} \left(
\frac{k_{\rm B} T}{2~{\rm keV}} \right)^{5/2} ~{\rm kpc}^2 ~ {\rm Myr}^{-1},
\nonumber
\end{eqnarray}
where $v^2_{\rm th,e} = 2 k_{\rm B} T / m_{\rm e}$ is the thermal speed of the 
electrons and $\tau_{\rm ee}$ is the electron--electron collision time \citep{cs04}.
The parallel thermal diffusivity of the ions is a factor $\sim$$( m_{\rm e} / m_{\rm i} )^{1/2}$ 
smaller \citep{spitzer62}. For future reference, we also define the (Spitzer) parallel 
thermal conductivity 
\begin{equation}
\chi = (p/T)\kappa.
\end{equation}

\subsubsection{Radiative cooling}

The second term on the right side of \eqref{eqn:pressure} is the radiative 
cooling rate function $\Lambda$. We 
adopt thermal Bremsstrahlung radiation, and take $\Lambda$ to be
\begin{equation}
\Lambda = 10^{-27} \left( \frac{ n_{\rm i} }{0.01~{\rm cm}^{-3}} \right)^2
\left( \frac{ k_{\rm B} T }{2~{\rm keV}} \right)^{1/2} ~{\rm ergs ~ cm}^{-3} ~
{\rm s}^{-1}
\end{equation}
\citep{rl79}. The cooling in the ICM is dominated by Bremsstrahlung above 
temperatures $\sim$$1~{\rm keV}$.

\subsection{Model geometry}\label{sec:geometry}

The stability calculation requires us to set up a 
suitable geometry and background magnetic field configuration. 
Galaxy clusters are approximately spherical, but in this geometry it is unclear 
what the most appropriate (or analytically feasible) configuration the 
magnetic field should take. Previous linear theory has sidestepped these issues 
by employing a local model, in which the magnetic field may take a simple form 
(purely vertical, for instance). However, in this paper we explicitly wish to 
test the importance of global structure when in the presence of Braginskii 
viscosity \citep{kunz11}. We hence employ an intermediate model, the 
plane-parallel atmosphere \citep[see e.g.][]{lamb97}, 
used frequently in stellar convection problems \citep{hpwb89,bjnrst96}. 
This is an idealised physical system, but 
one that can help isolate features associated with a cluster's global radial 
structure while evading the uncertainties and analytic difficulties of magnetic 
field configurations in spherical geometry. Though our model poorly describes a 
cluster in its full complexity, it should nevertheless exhibit its salient 
physics, shared by more advanced models.

The plane-parallel atmosphere is an approximation to a chunk of the 
spherical cluster. Terms arising from spherical geometry are dropped; angular 
structure is neglected but the radial structure is retained. Put another way, 
the model is local in the angular (i.e.\ horizontal, $x$, $y$) directions and 
global in the radial (i.e.\ vertical, $z$). 
We adopt a uniform gravitational acceleration in the vertical direction, 
$\bb{g} = -g \ez$. The vertical extent of the layer is taken to be finite and 
conditions need to be applied at its upper and lower boundaries, which occur at 
$z=0$ and $z=Z$. Finally, the background magnetic field is assumed to be 
constant and vertical within the layer. The plane-parallel atmosphere is 
well-suited to the 
morphology of the HBI modes in this set-up because these possess small radial 
and large horizontal wavenumbers \citep{kunz11}. 
The `quasi-global' plane-parallel atmosphere 
works best in the outskirts of the cool core: 
neglecting the geometric spherical terms introduce errors of order $Z/r$. 
However, these errors we believe will not alter the qualitative behaviour 
revealed by our calculations.

\section{Equilibrium state}\label{sec:equilibrium}

We consider a static Maxwellian plasma vertically-stratified in both density 
$\rho(z)$ and temperature $T(z)$, threaded by a uniform background magnetic 
field oriented along $\ez$. We further assume that the ratio of the thermal and 
magnetic pressure is large:
\begin{equation}
\beta \equiv \frac{8 \pi p}{B^2} = \frac{2 \vthsq}{\vasq} \gg 1 ,
\end{equation}
where $v_{\rm A} \equiv B / ( 4 \pi \rho )^{1/2}$ is the Alfv\'{e}n speed. 
Faraday rotation measurements suggest $\beta \sim 10^2$--$10^4$ in the cool 
cores of galaxy clusters \citep[for a review, see][]{ct02}.

The equilibrium must satisfy force and energy balances
\begin{equation}
\deriv{z}{p} = - g \rho \quad {\rm and} \quad \deriv{z}{q} = - \Lambda, 
\label{eqn:dynamicalequilibrium}
\end{equation}
where the heat-flux in the background state is given by
\begin{equation}\label{eqn:backgroundQ}
q = -  \chi \deriv{z}{T}.
\end{equation}
We consider two thermal equilibria: one in which there is no cooling, 
$\Lambda=0$, and hence no conductive heating; and one in which $\Lambda\neq 0$ 
and Bremsstralung radiation balances conductive heating.

Finally, thermal boundary conditions need to be applied at the upper and lower 
boundaries of the layer, at $z=0$ and $z=Z$. Various boundary conditions may be 
invoked, but we choose the simplest: we force the top and bottom layers to have 
fixed temperatures: $T(0)=T_0$ and $T(Z)=T_Z$.

\subsection{No cooling: $\Lambda = 0$}

In order to preserve thermal equilibrium when cooling is absent, the background 
heat-flux must be constant:
\begin{equation}\label{eqn:noheating}
\deriv{z}{} \left( \chi \deriv{z}{T} \right) = 0 .
\end{equation}
Enforcing the boundary conditions, equation (\ref{eqn:noheating}) may be 
integrated to yield the temperature profile
\begin{equation}
T(z) = T_0 \left( 1 + \zeta \frac{z}{Z} \right)^{2/7} ,
\end{equation}
where $\zeta = [( T_Z / T_0 )^{7/2} - 1]$ measures the magnitude of the constant 
heat-flux through the atmosphere ($q \propto - \zeta$). Combining this 
result with equation (\ref{eqn:dynamicalequilibrium}) determines the pressure profile
\begin{equation}
p(z) = p_0 \exp \left\{ - \frac{7}{5} \frac{Z}{\zeta H_0} \left[ \left( 1 +
\zeta \frac{z}{Z} \right)^{5/7} - 1 \right] \right\} ,
\end{equation}
where
\begin{equation}\label{eqn:H-defn}
H_0 = v^2_{\rm th,0} / g
\end{equation}
is the thermal-pressure scale height at $z = 0$, and $p_0$ is the pressure at 
$z=0$. The density follows from the equation of state.

This equilibrium is characterized by two dimensionless parameters: $\zeta$, 
which is $\sim$$10$--$50$ in typical cool-core clusters, and
\begin{equation}
\mc{G} \equiv \frac{Z}{H_0}
\end{equation}
which is a measure of the vertical extent of the layer. One example of this 
atmosphere with $T_Z / T_0 = 2.5$ (i.e. $\zeta \simeq 23.7$) and $\mc{G}=2$ is 
shown in Figure \ref{fig:atmosphere} as the red lines.

\subsection{Bremsstrahlung cooling: $\Lambda \propto \rho^2 T^{1/2}$}

When Bremsstrahlung cooling is present, no analytic solution exists. Instead, 
we employ a shooting method to solve equations 
(\ref{eqn:dynamicalequilibrium})--(\ref{eqn:backgroundQ}), subject to three 
boundary conditions: $T= T_0$ and $\rho = \rho_0$ at $z=0$, and $T = T_Z$ at 
$z=Z$. The presence of cooling introduces another dimensionless free parameter,
\begin{eqnarray}
\lefteqn{
\mc{S} \equiv \frac{ (\Lambda / p )_0 }{ \kappa_0 / Z^2 } \simeq  18.4 \left(
\frac{Z}{250~{\rm kpc}} \right)^2 \left( \frac{n_{\rm i,0}}{0.01~{\rm cm}^{-3}}
\right)^2 \left( \frac{k_{\rm B} T_0}{2~{\rm keV}} \right)^{-3}  ,
}\nonumber\\*
\end{eqnarray}
which is the ratio of the radiative diffusion time (across a distance $Z$) and 
the characteristic cooling time (each evaluated at $z=0$). $\mc{S}$ is related 
to the Stefan number (Sf) used in some convection studies. If we scale space by 
$Z$, $\rho$ by $\rho_0$, and $T$ by $T_0$, the equations can be written in the 
following form, suitable for numerical work:
\begin{equation}
\deriv{z}{q} = - \mc{S} \rho^2 T^{1/2} ,
\end{equation}
\begin{equation}
\deriv{z}{T} = -q T^{-5/2} ,
\end{equation}
\begin{equation}
\deriv{z}{\rho} = -\mc{G} \rho T^{-1} + \rho q T^{-7/2} .
\end{equation}
The boundary conditions are hence $\rho(0)=1$, $T(0)=1$ and $T(Z)=(T_Z/T_0)$. 
Note that in the limit of small $\mc{S}$ (ineffective radiative cooling), the 
first equation gives a constant $q$ to leading order and we recover the 
`no-cooling' solution of the previous subsection.

\begin{figure}
\centering
\includegraphics[width=2.5in]{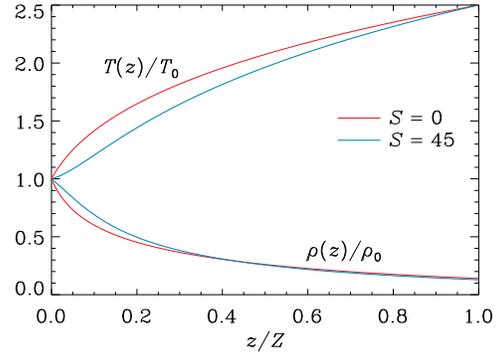}
\caption{Equilibrium atmospheres with (blue lines) and without (red lines)
Bremsstrahlung cooling for $T_Z / T_0 = 2.5$ (i.e. $\zeta \simeq 23.7$) and
$\mc{G} = 2$.}
\label{fig:atmosphere}
\end{figure}

Figure \ref{fig:atmosphere} describes an example of a 
cooling atmosphere with $T_Z / T_0 = 2.5$ (i.e. $\zeta \simeq 23.7$), 
$\mc{G}=2$, and $\mc{S}=45$, plotted with the blue lines. 
This atmosphere is very 
similar to the cool core of A85, particularly for $z \gtrsim 50~{\rm kpc}$ \citep{cdvs09}. 
Note that the effects of cooling are localised to relatively 
small $z$, where the density is greatest. As a result, the heat conduction $q$ 
(not plotted) is attenuated at these $z$, dropping to approximately zero at 
$z=0$. The gradient of the temperature follows, via \eqref{eqn:backgroundQ}, and 
is hence flatter than in the non-cooling atmosphere.

%
%

\section{Linear modes}\label{sec:linearshit}

\subsection{Linearised equations}\label{sec:lineareqns}

We consider two-dimensional perturbations, $\delta \rho$, $\delta T$, $\delta\BB$, 
and $\delta \bb{v}$, upon our equilibrium state that exhibit a space-time 
dependence $\propto f(z) \exp(\sigma t + \imag k x )$, where $\sigma$ is a 
(complex) growth rate and $k$ a (real) wavenumber. Such modes are local in the 
direction perpendicular to the background magnetic field ($\ex$) and 
global in the direction along the background magnetic field ($\ez$). 
Because the modes are two-dimensional, it is convenient to use the magnetic 
flux function $A$ defined through $\BB=\grad\btimes(A\,\ey)$. Finally, we scale 
space by $Z$, speed by $v_{\rm th,0}$, density by $\rho_0$, temperature by 
$T_0$, and the flux function by $B_0 Z$. Time, as a consequence, is scaled by 
$Z / v_{\rm th,0}$.

The governing equations, written to linear order in the perturbation 
amplitudes, are then
\begin{equation}\label{lineqn:continuity}
\sigma \frac{\delta\rho}{\rho} = - \imag k \delta v_x - \left( \deriv{z}{\ln\rho} 
+ \D{z}{} \right) \delta v_z ,
\end{equation}
\begin{eqnarray}
\lefteqn{
\sigma \delta v_x = - \imag k T \left( \frac{\delta\rho}{\rho} + \frac{\delta
T}{T} \right) +\frac{2}{\beta_0 \rho} \left( k^2 - \DD{z}{} \right) \delta A
} \nonumber \\* && \mbox{} 
- \frac{\imag k}{{\rm Re}} \frac{T^{5/2}}{\rho} \left( \frac{2}{3}
\D{z}{\delta v_z} -  \frac{1}{3} \imag k \delta v_x \right) ,
\label{lineqn:momx}
\end{eqnarray}
\begin{eqnarray}
\lefteqn{
\sigma \delta v_z = - T \D{z}{} \frac{\delta \rho}{\rho} - \left( T
\deriv{z}{\ln p} + T\D{z}{} \right)\frac{ \delta T}{T} 
} \nonumber \\* && \mbox{}
+ \frac{2}{{\rm Re}} \frac{T^{5/2}}{\rho} \left( \frac{5}{2}
\deriv{z}{\ln T} + \D{z}{} \right) \left( \frac{2}{3} \D{z}{\delta v_z} -
\frac{1}{3} \imag k \delta v_x \right) , 
\label{lineqn:momz}
\end{eqnarray}
\begin{eqnarray}
\lefteqn{
\sigma \frac{3}{2} \frac{\delta T}{T} = - \imag k  \delta v_x - \left(
\frac{3}{2} \deriv{z}{\ln T} + \D{z}{} \right) \delta v_z
}\nonumber \\* && \mbox{}
+ \frac{1}{p\,\text{Pe}} \left[ \DD{z}{}\left(T^{7/2}\frac{\delta
T}{T}\right) + \imag k q \D{z}{\delta A} - \mc{S}\,\delta\Lambda \right],
\label{lineqn:pressure}
\end{eqnarray}
\begin{equation}
\sigma \delta A = -\delta v_x ,
\label{lineqn:induction}
\end{equation}
where
\begin{equation}
\delta \Lambda = \Lambda\left( 2\frac{\delta\rho}{\rho} + \frac{1}{2}
\frac{\delta T}{T}\right) .
\end{equation}
These equations introduce three dimensionless parameters: the plasma beta 
at $z=0$, denoted by $\beta_0$, and the Reynolds and Peclet numbers at $z=0$, 
defined through
\begin{equation}
{\rm Re} = \frac{Z v_{\rm th,0}}{\nu_0} \quad {\rm and} \quad {\rm Pe} =
\frac{Z v_{\rm th,0}}{\kappa_0},
\end{equation}
which quantify the relative importance of viscous and thermal transport, 
respectively. In the context of the ICM, both numbers can be replaced by a 
single parameter, the inverse Knudsen number at $z=0$, defined through 
\begin{align}
{\rm Kn}_0^{-1} &\equiv \frac{H_0}{\lambda_{{\rm mfp},0}} \\
 & \simeq 1207 \left( \frac{g}{10^{-8} \,\text{cm}\,\text{s}^{-2}} \right)^{-1}
\left( \frac{n_{\rm i,0}}{0.01~{\rm cm}^{-3}} \right) \left( \frac{k_{\rm B}
T_0}{2~{\rm keV}} \right)^{-1}  .\nonumber
\end{align}
The Knudsen number is simply the ratio of the mean free path to the scale 
height. We find that
\begin{equation}
{\rm Re} = 2.08\,\mc{G} {\rm Kn}^{-1}_0 \quad {\rm and} \quad {\rm Pe} =
0.042\,\mc{G} {\rm Kn}^{-1}_0.
\end{equation}
There are hence six dimensionless parameters that fully specify the problem. 
The equilibrium is set by $\xi$, $\mc{G}$, and $\mc{S}$, while the linear 
analysis introduces $\beta_0$, ${\rm Kn}_0$, and the (scaled) horizontal 
wavenumber $k$.

Finally, the Prandtl number may be written as
\begin{equation}
{\rm Pr} \equiv \frac{\nu_0}{\kappa_0} = 0.607 \, \frac{\Lambda_{\rm
e}}{\Lambda_{\rm i}} \left( \frac{2 m_{\rm e}}{m_{\rm i}} \right)^{1/2} \simeq
0.02 .
\end{equation}
Though Pr does not appear explicitly in the governing equations it is still of 
interest. Note that ${\rm Pr}$ is roughly constant. This implies that viscous forces 
operate on a timescale $\tau_{\rm visc}$ that is a fixed number larger than the 
timescale on which conduction operates, $\tau_{\rm cond}$. For the approximately 
incompressible perturbations examined in this paper, $\tau_{\rm visc} / 
\tau_{\rm cond} = (2/15) {\rm Pr}^{-1} \approx 6$ \citep{kunz11}.

\subsubsection{Boundary conditions}\label{sec:boundaries}

This set of coupled ordinary differential equations must be solved over the 
domain $z=[\,0,1\,]$ subject to boundary conditions. 
We first assume that there 
is no penetration through the upper and lower boundaries of the atmosphere, and 
so $\delta v_z = 0$ at $z=0$ and $1$. In addition, the magnetic field is 
constrained to be purely vertical at the boundaries: $ {\rm d} \delta A / {\rm d} z = 0$ 
at $z=0,\,1$. Now, if we enforce constant temperatures $T_0$ and $T_Z$ 
at $z=0,\,1$, then we must also have zero temperature fluctuations at the 
boundary, $\delta T=0$. This boundary condition is used in most numerical 
studies of the HBI in plane-parallel geometry and is the simplest to implement.
Other conditions were trialed, such as zero heat-flux perturbation on the lower boundary, 
with little change in the qualitative results.

\subsection{Numerical scheme}\label{sec:numerics}

We solve equations (\ref{lineqn:continuity})--(\ref{lineqn:momz}) numerically 
via a pseudospectral technique. The $z$-domain is partitioned into a 
Gauss-Lobatto grid with $N = 250$ grid points and the differential operator is 
discretized as a (Chebyshev) derivative matrix (Boyd 2000). A $5N \times 5N$ 
matrix equation ensues taking the form of a generalised algebraic eigenvalue problem; 
this yields a spectrum that approximates that of the original differential 
operator. The eigenvalues $\sigma$ are obtained using standard linear algebra 
routines such as the QZ algorithm or an Arnoldi method (Golub \& van Loan 1996).

%
%
\begin{figure*}
\centering
\includegraphics[height=2.5in]{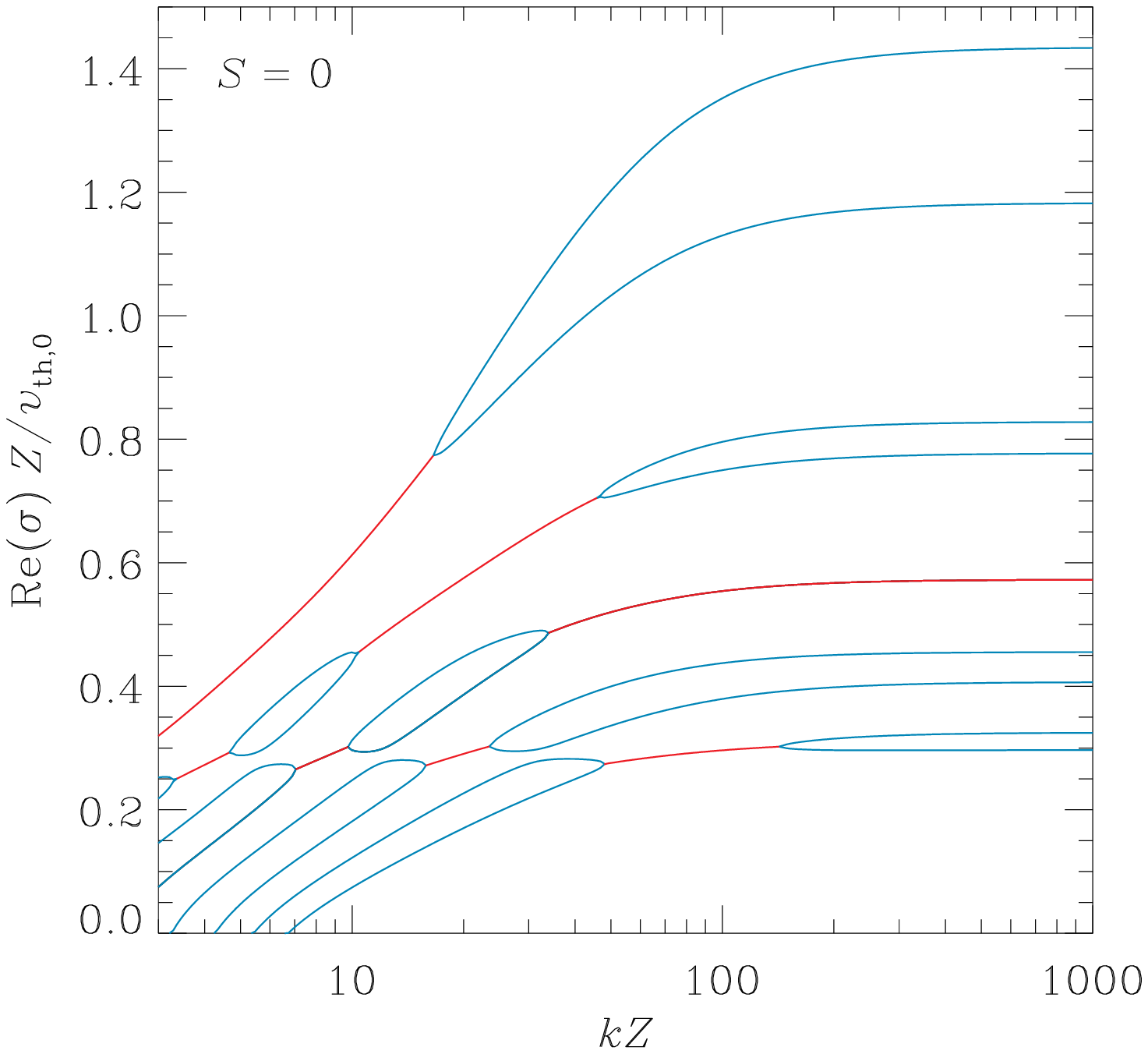}
\qquad
\includegraphics[height=2.5in]{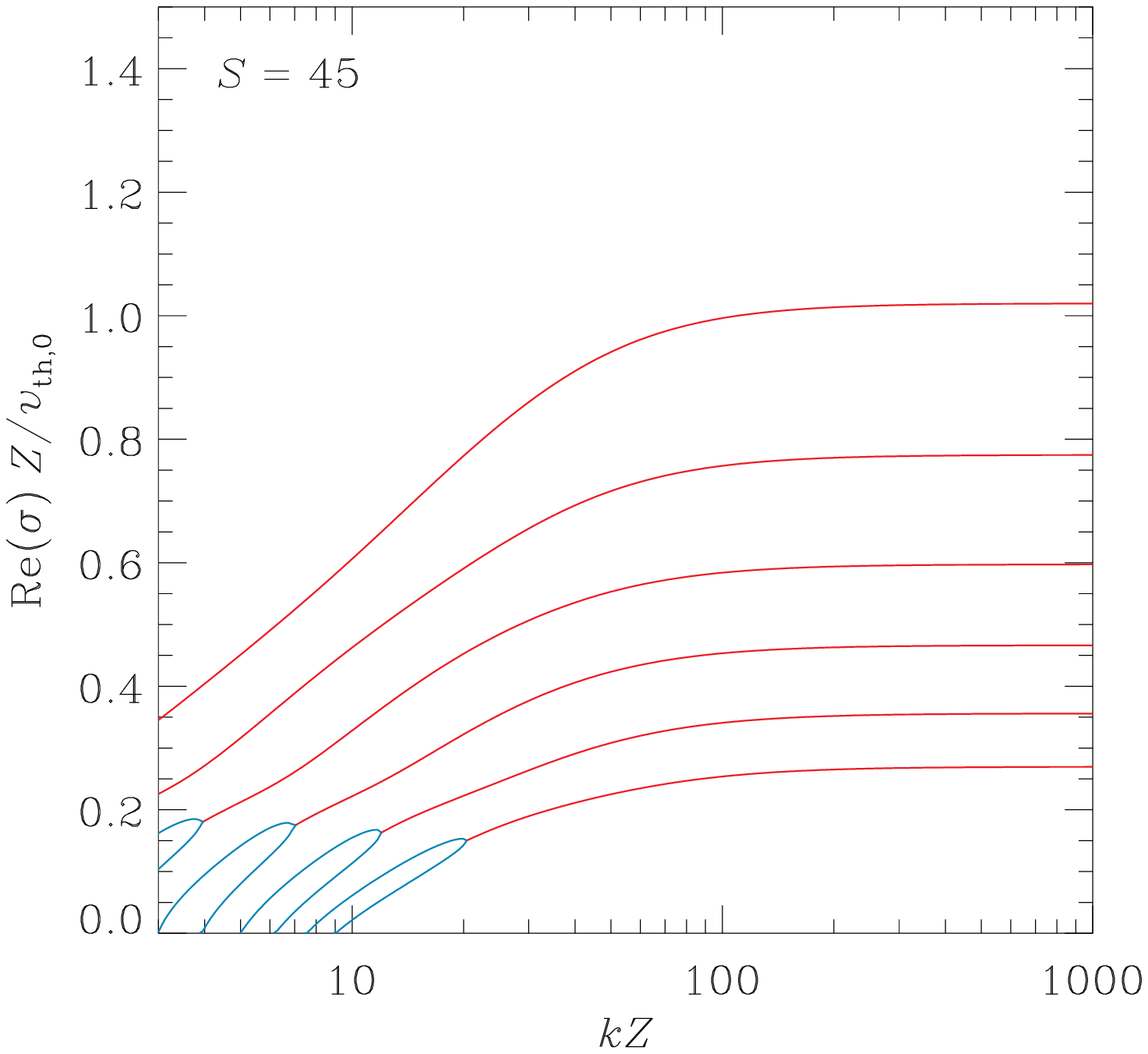}
\caption{Growth rates of various HBI modes as functions of the 
perpendicular wavenumber $k$. The parameters of the system are: $T_Z/T_0=2.5$, 
$\mc{G}=2$, $\beta_0=10^5$, and ${\rm Kn}^{-1}_0 = 1500$. The left panel 
corresponds to the non-cooling case $\mc{S}=0$, the right panel to $\mc{S}=45$. 
The growth rates are either purely real (blue lines) or come in complex 
conjugate pairs (red lines). Note that there exist multiple bifurcations at 
which conjugate pairs `detach' and become purely monotonic modes, or two 
monotonic modes coalesce and form a conjugate pair. The cooling case appears to 
favour oscillatory growth (complex conjugate pairs), especially for large $k$.} 
\label{fig:modes}
\end{figure*}

\subsection{Results}\label{sec:results}

We present results for various parameters. Throughout, however, we set $\mc{G} = 
2$, $T_Z/T_0=2.5$ (or equivalently $\zeta = 23.7$). This leaves the cooling 
parameter $\mc{S}$, which we either assign to be 0 or 45; the plasma beta at 
the lower boundary $\beta_0 = 10^5$; and the inverse Knudsen number 
${\rm Kn}_0^{-1}$, which we take usually to be 1500. Finally, the horizontal wavenumber 
$k$ ranges between 0 and 1000.

\subsubsection{Growth rates}\label{sec:growthrates}

On account of the finite vertical domain of the layer, there are a discrete 
number of HBI modes each with a distinct vertical structure, as opposed to the 
$k_z$ continuum exhibited by the local model. The fastest growing of these 
modes undergo the least vertical variation (corresponding roughly to smaller 
$k_z$ in the local analysis), while the slowest growing modes exhibit much 
finer-scale structure (larger $k_z$).

In Fig.~\ref{fig:modes} the growth rates of the leading modes are plotted 
against horizontal wavenumber $k$ for the non-cooling and cooling cases. In 
accord with the local analysis, we find that the fastest-growing HBI modes have 
large horizontal (perpendicular) wavenumber, with 
$k Z \gtrsim 3 \mc{G} {\rm Kn}^{-1/2}_0$ \citep[see figure 3 of][]{kunz11}. 
This wave configuration maximizes 
convergence/divergence of the background heat-flux and the consequent 
heating/cooling of the plasma. Modes on arbitrarily small horizontal scales 
grow at the maximum rate, and so the system is ill-posed unless these small 
scales are regularised via the inclusion of finite Larmor radius (FLR) effects, 
for example. Conversely, most modes are extinguished for 
sufficiently small $k$ (long horizontal wavelengths): 
an HBI mode ceases to grow efficiently when its 
horizontal wavelength becomes much longer than its characteristic vertical 
scale. The concentration of the heat-flux becomes inefficient in this case.

The magnitude of the growth rates that appear in Fig.~\ref{fig:modes} is 
typical for sensible parameter choices. The peak e-folding times are roughly 
$Z / v_{\rm th,0} \approx 400 ~{\rm Myr}$ for $Z = 250~{\rm kpc}$ and 
$k_{\rm B} T_0 = 2~{\rm keV}$. As $\beta_0$ is lowered, slower-growing modes, which 
exhibit finer vertical structure, are stabilised because the Alfv\'en length 
reaches their characteristic lengthscales and magnetic tension comes into play. 
For instance, when $kZ=300$ and $\beta_0=10^5$ (with no cooling) there are 32 
growing modes; but when $\beta_0$ is decreased to $10^4$ this number drops to 
$14$, and then to $5$ when $\beta_0=10^3$. Greater magnetic tension (lower 
$\beta_0$) also lowers the maximum growth rate.

Unstable HBI modes can exhibit either monotonic or oscillatory growth 
analogous to magnetoconvection in a finite layer \citep[e.g.][]{pw82}. 
In the case of oscillatory growth, modes occur in complex 
conjugate pairs. As $k$ (or $\text{Kn}_0$) varies, bifurcations occur where 
conjugate pairs `detach' and become purely monotonic modes, or monotonic 
modes coalesce and transform into conjugate pairs. In Fig.~\ref{fig:modes} 
these two modes are represented by either blue (purely real) or red (complex) 
lines. This behaviour should be contrasted to the local 
analysis in which the HBI is always monotonically growing. It follows, in part, 
from the variation in the background temperature and, in particular, its 
influence on the thermal conductivity. In a global model the perturbed heat 
flux introduces a term proportional to $(dT/dz)\delta\chi\propto (dT/dz) \delta T$, 
which is absent from a local analysis. Under the right conditions, this 
term can facilitate growth by injecting the free energy from the gradient into 
a thermodynamic wave during its `compressed' phase. Interestingly, 
in a radiatively cooling atmosphere growing modes tend to 
favour the oscillatory form. Cooling accentuates local 
perturbations in $\delta T$ and thus reinforces the oscillatory response of the 
perturbed heat-flux.

%
%
\begin{figure*}
\centering
\includegraphics[angle=90,height=2.8in]{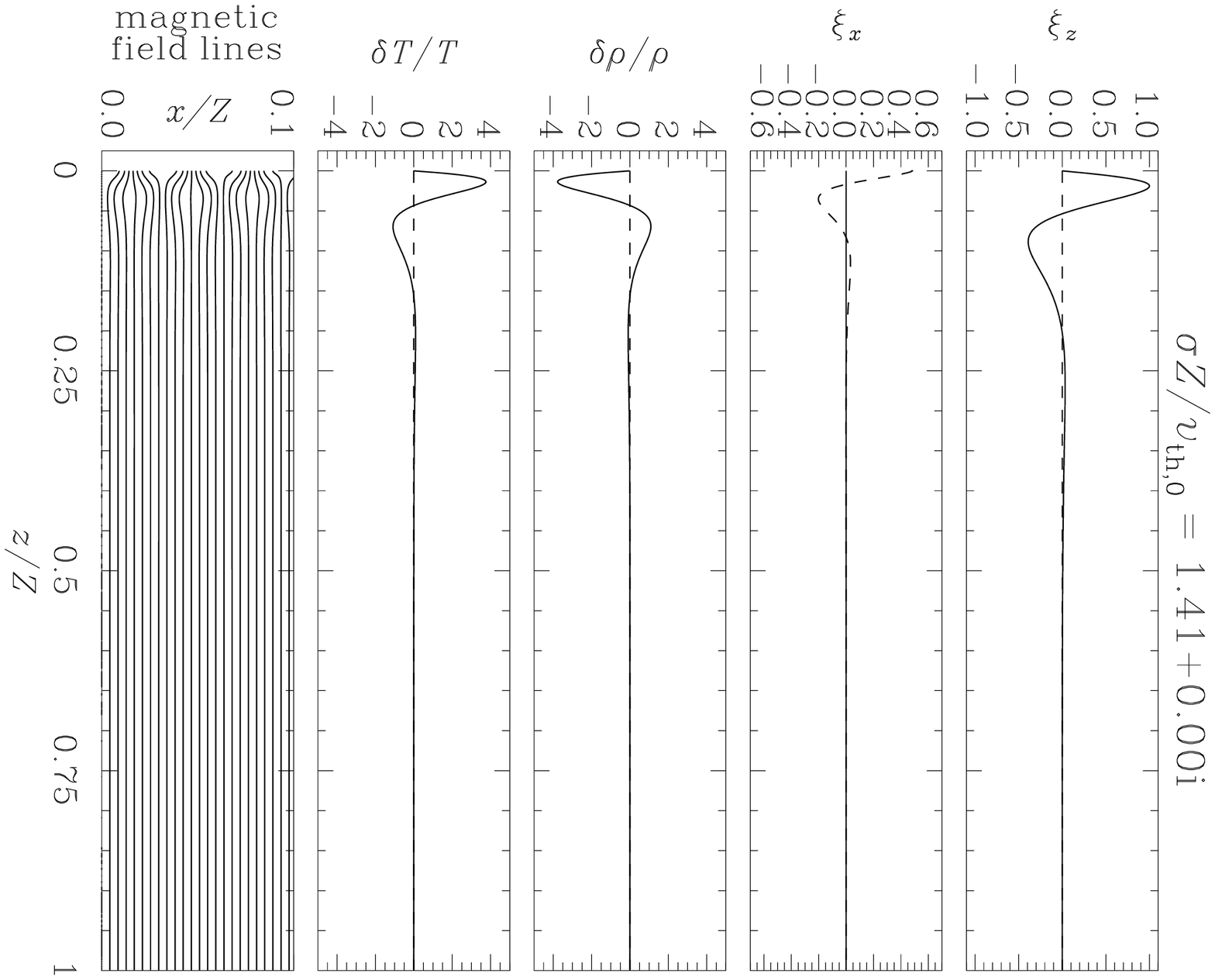}
\qquad\qquad
\includegraphics[angle=90,height=2.8in]{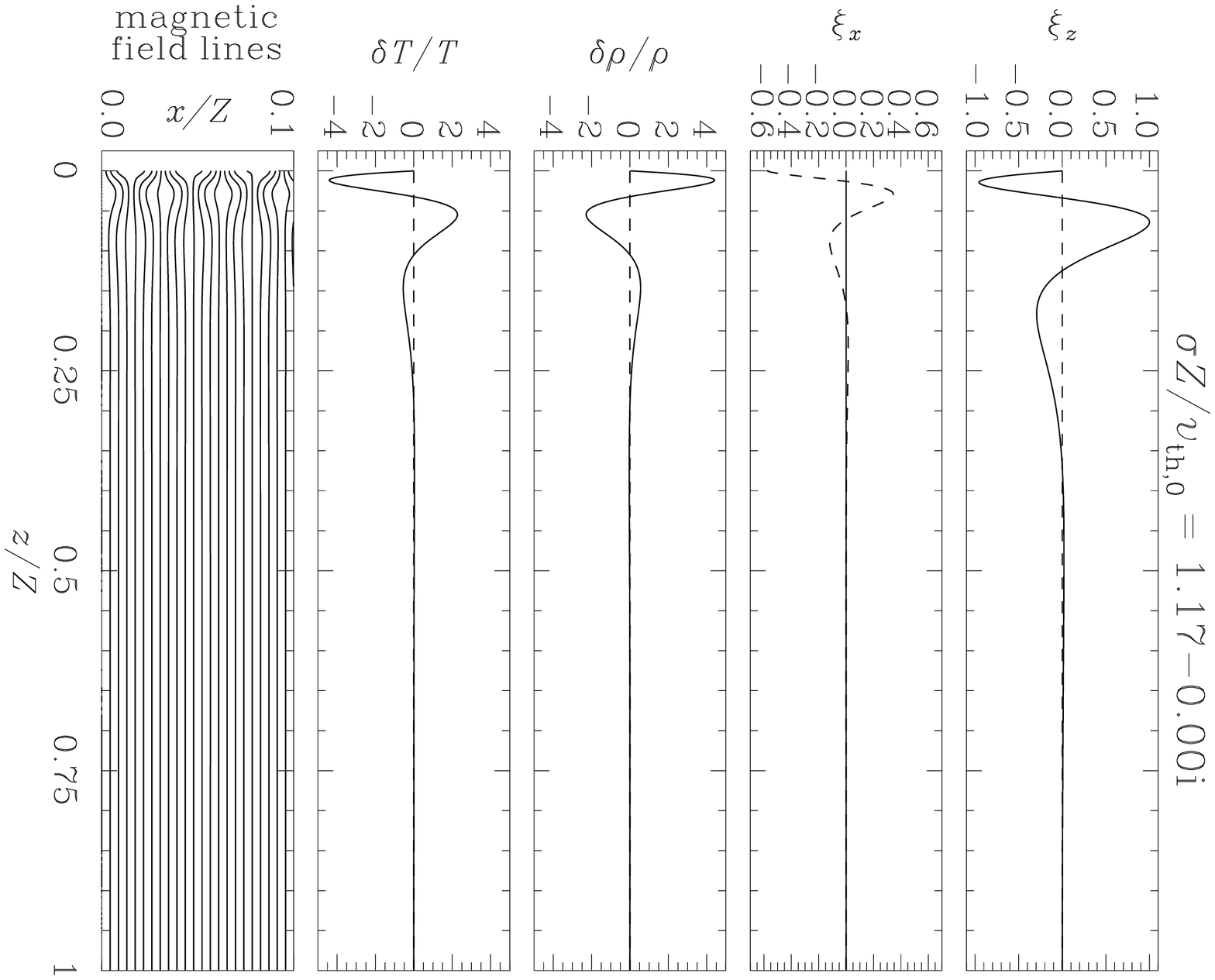}
\newline\newline
\includegraphics[angle=90,height=2.8in]{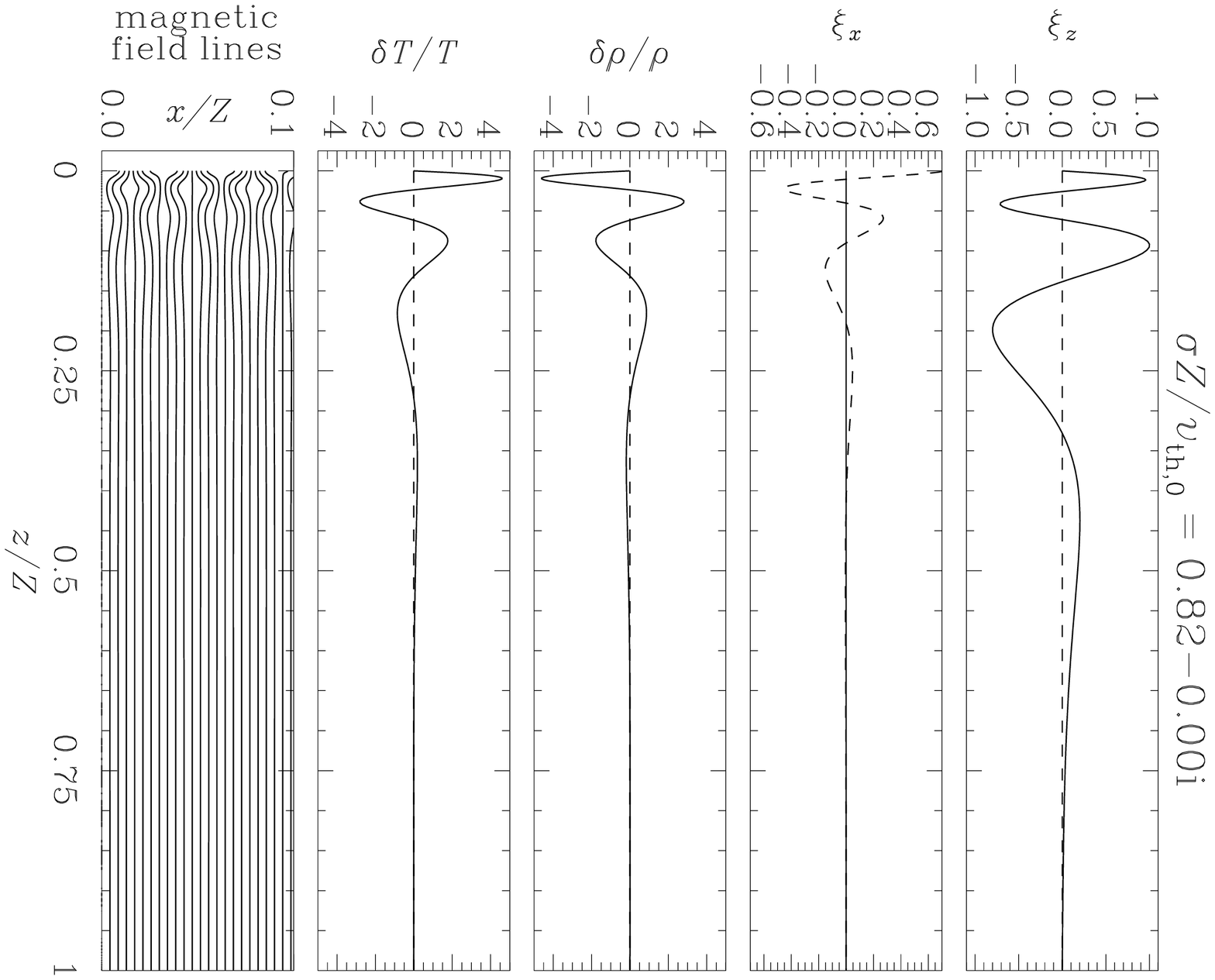}
\qquad\qquad
\includegraphics[angle=90,height=2.8in]{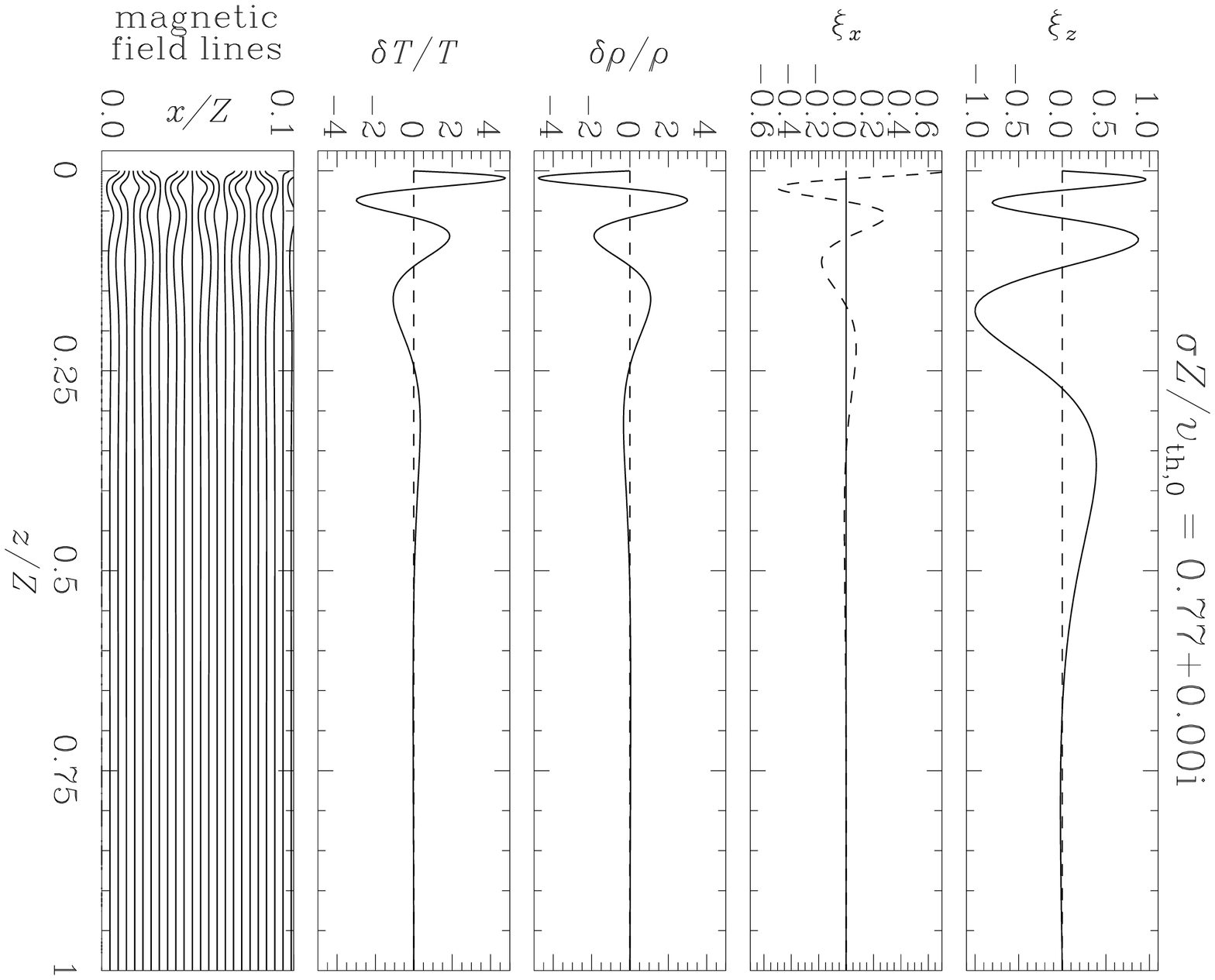}
\newline
\caption{Eigenvectors ($\xi_x = \delta v_x / \sigma$, $\xi_z = \delta v_z / 
\sigma$, $\delta\rho / \rho$, $\delta T / T$) and magnetic field lines of the 
four fastest-growing modes for $\mc{S}=0$, ${\rm Kn}^{-1}_0 = 1500$, $\beta_0 = 
10^5$, and the horizontal wavenumber $kZ=250$. The real (imaginary) part of 
each eigenvector is denoted by the solid (dashed) line. The growth rates are 
given at the top of each plot; for reference, $Z / v_{\rm th,0} \approx 400 
~{\rm Myr}$ for $Z = 250~{\rm kpc}$ and $k_{\rm B} T_0 = 2~{\rm keV}$.}
\label{fig:nocool}
\end{figure*}
%
%
%

%
%
\begin{figure*}
\centering
\includegraphics[angle=90,height=2.8in]{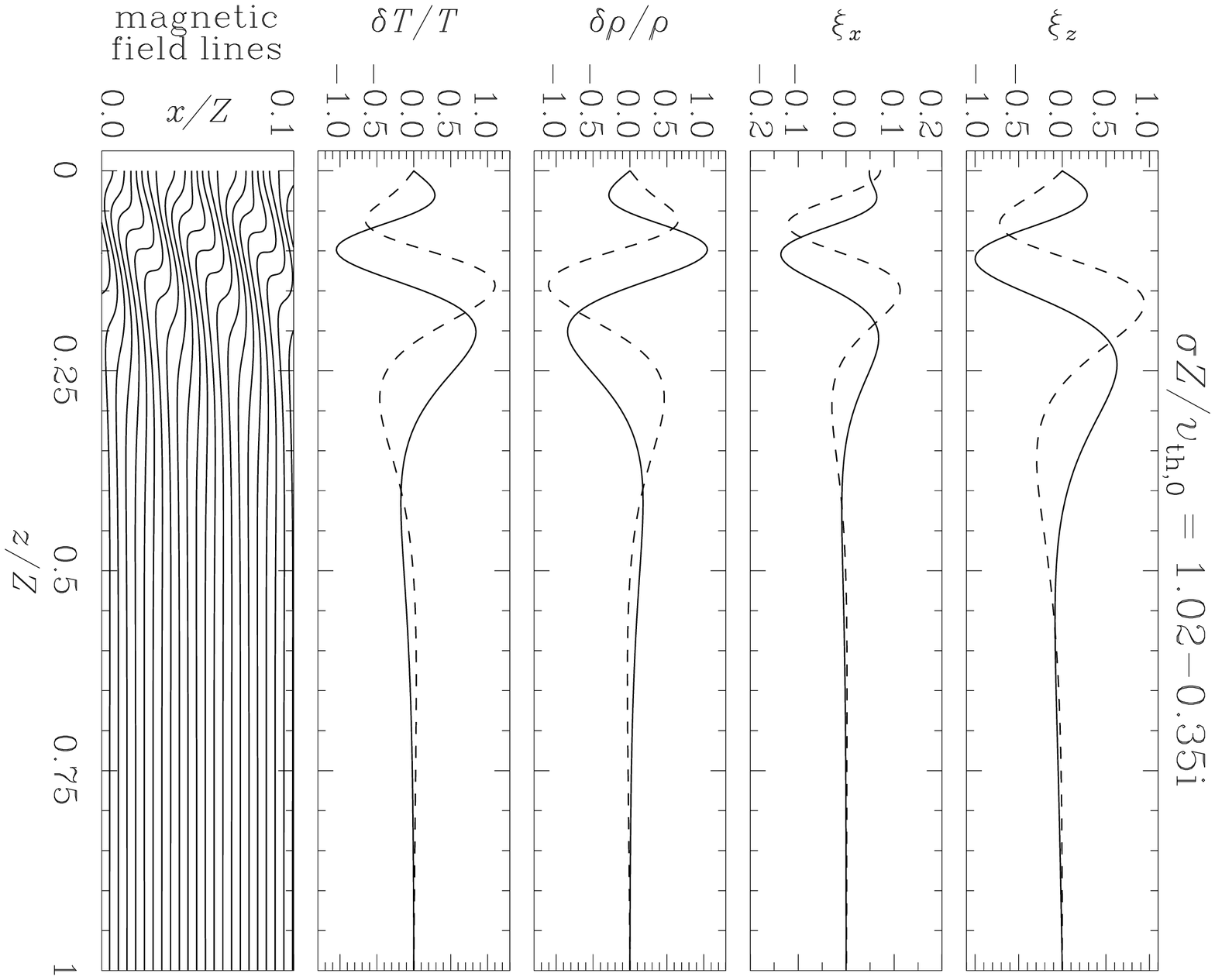}
\qquad\qquad
\includegraphics[angle=90,height=2.8in]{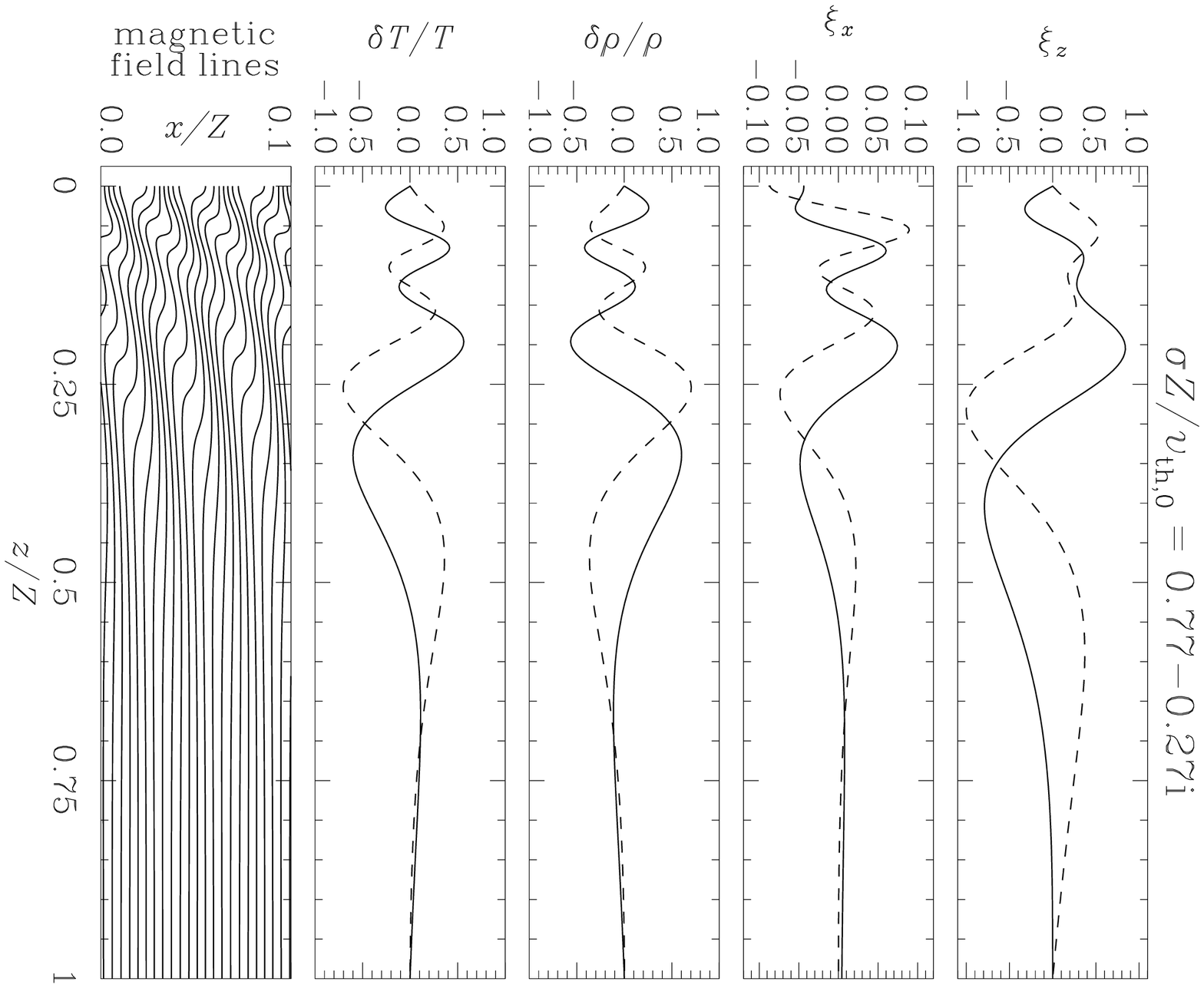}
\newline\newline
\includegraphics[angle=90,height=2.8in]{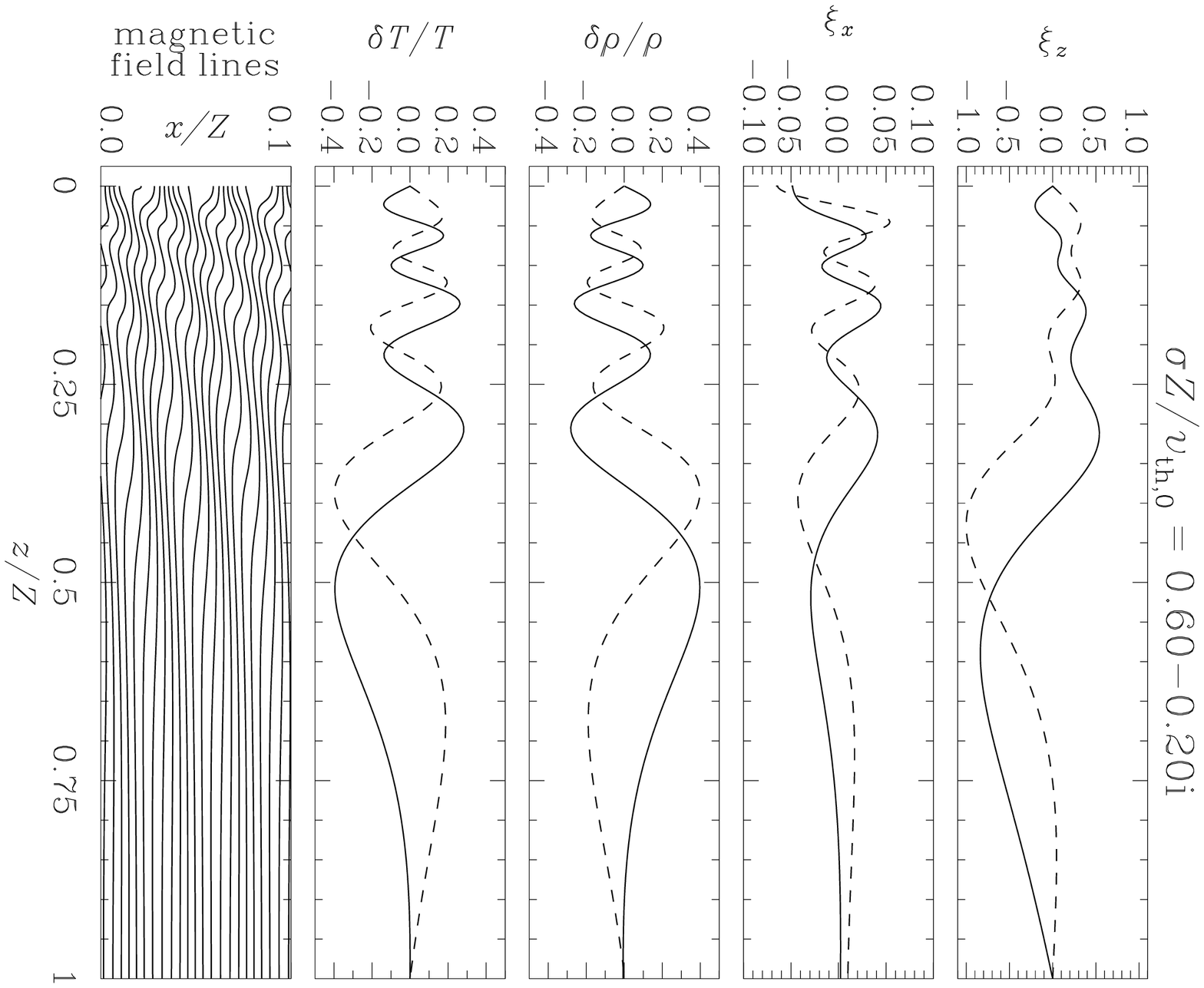}
\qquad\qquad
\includegraphics[angle=90,height=2.8in]{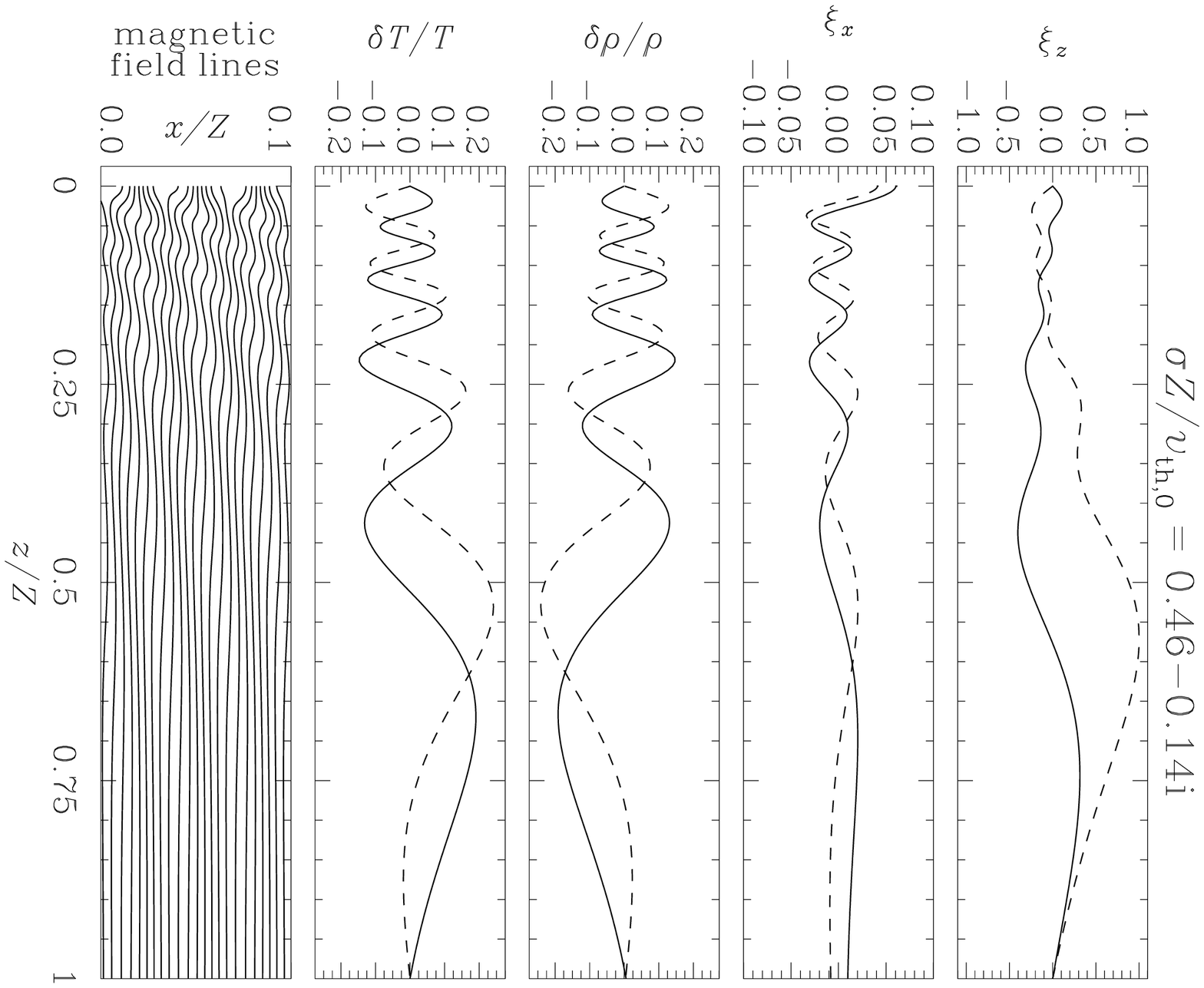}
\newline
\caption{Eigenvectors ($\xi_x = \delta v_x / \sigma$, $\xi_z = \delta v_z / 
\sigma$, $\delta\rho / \rho$, $\delta T / T$) and magnetic field lines of the 
four fastest-growing modes for a cooling atmosphere. Parameters are the same as 
in Fig.~\ref{fig:nocool} except that $\mc{S}=45$. These modes occur in complex 
conjugate pairs, but we plot only one member of each pair.}
\label{fig:cool}
\end{figure*}

\subsubsection{Eigenfunctions}

In Fig.~\ref{fig:nocool} we show the four fastest-growing eigenmodes for 
$\mc{S} = 0$, ${\rm Kn}_0^{-1} = 1500$, and $\beta_0 = 10^5$. The horizontal 
wavenumber $k=250$. We plot their Lagrangian displacements, $\xi_x$ and 
$\xi_z$, the scaled density $\delta \rho/\rho$ and temperature $\delta T/T$, as well 
as the perturbed magnetic-field lines. 

What is most striking in these plots is that the the fastest-growing modes, 
which should dominate the driving of nonlinear disordered flows, are confined 
to the innermost $\sim$$20\%$ of the atmosphere, as presaged by the local WKBJ approach 
(Kunz 2011). There are two reasons for 
this. First, because the HBI growth rate is proportional to $({\rm d}\ln T / 
{\rm d}\ln z)^{1/2}$, these modes take advantage of the steeper temperature 
gradient at small $z$. Indeed, the local growth rate is a factor 
$\zeta^{1/2}\approx 5$ less at the top of the layer than it is at the bottom. 
Second, because the kinematic viscosity is $\nu \propto 
T^{5/2}/\rho$, the higher temperatures and lower densities characteristic of 
the upper portion of the atmosphere guarantee rapid viscous damping of 
wavelengths significantly smaller than the local thermal-pressure scale height. 
This is also the cause of the relatively straight field lines for $z \gtrsim 0.2 Z$ 
for all of the shown eigenmodes. As a consequence of these two effects, the 
degree of confinement depends on the parameters $(T_Z/T_0)$ and ${\rm Kn}_0$. 
We find that the fraction of the atmosphere perturbed by the fastest growing mode scales like 
$(T_Z/T_0)^{-5/2} {\rm Kn}^{1/2}_0$ for ${\rm Kn}_0 > 1/3000$. The 
dependence on ${\rm Kn}_0$ here coincides, unsurprisingly, with 
the HBI's favoured vertical wavelength in the local analysis \citep{kunz11}.

Slower growing modes (not shown) exhibit greater variation (i.e. more nodes) and 
extend farther into the atmosphere. Yet only the very slowest modes exhibit significant 
perturbations throughout the entire layer. For example, a typical e-folding time for the 
growth of substantial magnetic structures at large $z$ is $\gtrsim$$5~{\rm Gyr}$, an 
order of magnitude longer than the peak e-folding time. Consequently, it is unlikely 
that such modes are dynamically important during the nonlinear saturation of the HBI.

Figure \ref{fig:cool} presents the four fastest-growing eigenmodes for the same 
parameters as in Fig.~\ref{fig:nocool}, except that $\mc{S} = 45$. 
These modes occur in complex conjugate 
pairs, but we plot only one member of the pair. The modes' 
morphologies are qualitatively the same as in the non-cooling case; however the 
localisation is less pronounced. This is due in part because the equilibrium 
temperature gradient and heat-flux are smaller at low $z$ when cooling is 
present. Thus both the free energy source and the catalyst for instability are 
diminished. To compensate, the mode `spreads out' to access as much 
energy as viscous damping permits. Nonetheless, maximum HBI growth rates suffer 
and drop, in relative terms, by roughly a third (cf.\ Fig.~\ref{fig:modes}).  
As with the non-cooling atmosphere, the growth of significant magnetic structures 
at large $z$ is relatively slow, with e-folding times of at least $5~{\rm Gyr}$.

\subsection{Comparison with simulations}

Other than elucidating the nature of the HBI under global cluster conditions, 
the above analysis also serves as a useful testbed for numerical codes that include 
anisotropic conduction and viscosity. In this subsection we offer a comparison of 
non-radiative HBI growth rates computed from the linear theory (Section 
\ref{sec:growthrates}) and from simulations using the Godonov code Athena 
\citep{sgths08} equipped with a Braginskii stress and anisotropic heat conduction. 
Specifically, we use the simulation run `H2dBrag' presented in \S~4.3 of \citet{kbrs12}. 

The simulation was initialized with an atmosphere identical to that denoted by the 
red lines in Fig. \ref{fig:atmosphere} with parameters 
the same as in Fig. \ref{fig:nocool}, i.e. $\beta_0 = 10^5$ and ${\rm Kn}^{-1}_0 = 1500$. 
Likewise, the boundary conditions on the top and bottom of the computational 
domain, which employed 512$\times$1024 zones to span $H_0 \times 2 H_0$, were 
the same as those described in Section \ref{sec:boundaries}. The equilibrium 
state was broken by seeding the velocity with small-amplitude white noise. Further 
details regarding this simulation may be found in \S\S~3--4 of \citet{kbrs12}.

\begin{figure}
\centering
\includegraphics[width=2.5in]{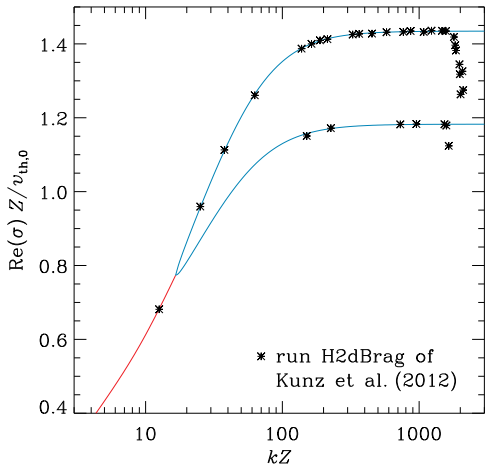}
\caption{Comparison between the analytical dispersion relation (red and blue
lines; as in Fig. 
\ref{fig:nocool}) and the numerical dispersion relation (asterisks; obtained
from run H2dBrag of \citep{kbrs12} for the two fastest-growing modes. 
The parameters of the system are: $T_Z / T_0 = 2.5$,
$\mc{G} = 2$, $\beta_0 = 10^5$, and ${\rm Kn}^{-1}_0 = 1500.$}
\label{fig:code}
\end{figure}

In Fig.~\ref{fig:code}, we superimpose the numerical dispersion relation 
(asterisks) on the analytical dispersion relation for the two fastest-growing 
modes. The numerical growth rates were obtained by Fourier transforming a 
periodic reconstruction of the simulation domain, selecting prominent and easily 
identifiable modes, and tracing their evolution throughout the linear phase of 
growth ($t v_{\rm th,0} / Z \approx 2$--$5$). The agreement between the code 
results and the analytical prediction is extremely good, up to wavenumbers 
approaching those associated with the grid spacing ($k_{\rm max} = 1024\pi$). 
The corresponding real-space profiles of the temperature and density fluctuations 
(not shown here) are also in remarkable agreement with those predicted in 
Fig.~\ref{fig:nocool}. This is a valuable check and cross-validates both the linear 
theory and the numerical code.

Because of numerical dissipation on the grid, at $k \gtrsim k_{\rm max} / 2$ the 
numerical growth rates abruptly depart from the theoretical prediction and 
drop precipitously. At which $k$ this `unphysical' regularisation occurs 
depends entirely on the resolution adopted, and formally 
such simulations can never be resolved. In practice, however, 
it is unclear how many of these small-scale HBI modes are actually 
required to adequately describe the nonlinear dynamics of the system. 
Also unclear is whether the grid regularisation realistically 
approximates regularisation by FLR effects, and if this matters in the 
saturated state. Some of these issues are addressed in \citet{kbrs12}.

%
%

\section{Discussion}\label{sec:discussion}

In this paper we have examined the linear stability of weakly collisional, 
thermally stratified atmospheres 
in which the conduction of heat and the viscous dissipation of motions are 
anisotropic with respect to the 
magnetic field. Such atmospheres are representative of those found in the 
cooling cores of non-isothermal 
galaxy clusters, which are subject to a heat-flux buoyancy-driven instability 
(HBI).

Numerical simulations 
of the HBI have established that the instability acts in such a way as to 
disrupt the conductive flow of heat, 
ultimately insulating the core from the hot thermal bath at 
large radii. In the presence of 
appreciable radiative losses, a cooling flow inevitably develops unless 
turbulent stirring by other 
means can reopen the field lines and reinstate the conductive flux.

The pressure anisotropies generated by the HBI play an important role in 
suppressing the instability by 
reducing growth rates and by shifting the fastest-growing modes to relatively 
long wavelengths. In the bulk of 
cool-core clusters, the collisionality is sufficiently low that these 
wavelengths become comparable to the 
thermal-pressure scale height of the gas. The HBI then operates on a global 
scale, and the analysis we 
have performed here marks one contribution towards understanding how the HBI 
changes in a global setting.

Our main result is that the low collisionality beyond $\sim$$50~{\rm kpc}$ in 
typical cool-core clusters significantly 
impedes the otherwise disordered motions driven by the HBI. As a result, the 
magnetic-field lines can remain 
relatively straight over a significant fraction of the Hubble time and the 
conductive flow of heat may 
proceed at an appreciable fraction of the Spitzer value. Near-complete 
field-line insulation due to the HBI could very 
well occur, but our analysis suggests that it is well-localised to the 
innermost regions of the cool core. 
This highlights 
the need for radio-mode feedback at these scales from a powerful 
central dominant galaxy.

This conjecture is supported by recent numerical simulations of the 
HBI presented in \citet{kbrs12}, which 
self-consistently take into account the pressure anisotropy driven by the HBI 
and its adverse consequences for the 
instability's development. In contrast, 
\citet[][hereafter P12]{pmqs12} perform analogous simulations of 
the HBI but these show that 
pressure anisotropy has little effect on the nonlinear saturation of the instability; 
the HBI operates throughout the bulk of the cluster core unhampered and 
imposes significant field-line insulation throughout. 
These very different conclusions possibly arise from dissimilar parameters and models. 
In particular, the P12 simulations that 
employ a plane-parallel atmosphere are of very small vertical extent, typically 
$Z= 0.2H_0$. As a consequence, there is insufficient room for 
global effects to come into play. These simulations 
cannot exhibit the mode localisation that we emphasise. 
However, P12 also perform two spherical global simulations of 
the HBI in model cool-core clusters, and these support the 
results of their plane-parallel simulations. Here the contradiction with our work 
is more puzzling, but probably arises from 
different parameters. For instance, P12's ICM model is
far more collisional and less magnetised than ours, with
$\beta \sim 10^7$. The effective absence of magnetic tension in these simulations 
may be important as magnetic tension suppresses 
those shorter-scale (local) HBI modes that function throughout the core. 
Alternatively, the discrepancy may be due to P12's 
different initial magnetic-field geometry (tangled 
on scales $30$--$50~{\rm kpc}$) or due to a basic dissimilarity 
between the spherical and plane-parallel geometries. 
But the apparent disagreement is a spur for future work: 
further three-dimensional simulations of cool-core clusters under representative 
conditions are needed.

\section*{Acknowledgments}
The authors would like to thank the reviewer, Steve Balbus, for an insightful 
review that helped clarify a number of issues in the manuscript.
HNL acknowledges funding from STFC through grant ST/G002584/1. 
MWK was supported by STFC grant ST/F002505/2 during the early phases of this
work and is currently supported by NASA through Einstein Postdoctoral Fellowship 
Award Number PF1-120084 issued by the Chandra X-ray 
Observatory Center, which is operated by the Smithsonian Astrophysical 
Observatory for and on behalf of NASA under contract NAS8-03060. 
This work was supported in part by the Leverhulme Trust Network for Magnetised
Plasma Turbulence. The authors thank Alex Schekochihin for many beneficial 
conversations and for reading through an earlier draft of the manuscript.

\label{lastpage}


\begin{thebibliography}{99}

\bibitem[Balbus(2000)]{balbus00}
Balbus S.~A., 2000, ApJ, 534, 420

\bibitem[Balbus(2001)]{balbus01}
Balbus S.~A., 2001, ApJ, 562, 909

\bibitem[Balbus \& Reynolds(2008)]{br08}
Balbus S.~A., Reynolds C.~S., 2008, ApJ, 681, L65

\bibitem[Bogdanovi\'{c} et al.(2009)]{brbp09}
Bogdanovi\'{c} T., Reynolds C.~S., Balbus S.~A., Parrish I.~J., 2009, ApJ, 704,
211

\bibitem[Boyd(2000)]{boyd2000}
Boyd J.~P., 2000, Chebyshev and Fourier Spectral Methods (2nd ed.), Dover
Publications, New York.

\bibitem[Braginskii(1965)]{braginskii65}
Braginskii S.~I., 1965, Rev. Plasma Phys., 1, 205

\bibitem[Brandenburg et al.(1996)]{bjnrst96}
Brandenburg A., Jennings R.~L., Nordlund A., Rieutord M., Stein R.~F., Tuominen
I., 1996, JFM, 306, 325

\bibitem[Carilli \& Taylor(2002)]{ct02}
Carilli C.~L., Taylor G.~B., 2002, ARA\&A, 40, 319

\bibitem[Catto \& Simakov(2004)]{cs04}
Catto P.~J., Simakov A.~N., 2004, Phys. Plasmas, 11, 90

\bibitem[Cavagnolo et al.(2009)]{cdvs09}
Cavagnolo K.~W., Donahue M., Voit G.~M., Sun M., 2009, ApJS, 182, 12

\bibitem[Chew, Goldberger \& Low(1956)]{cgl56}
Chew C.~F., Goldberger M.~L., Low F.~E., 1956, Proc. R. Soc. London A, 236, 112

\bibitem[Golub \& van Loan(1996)]{gvl96}
Golub G.~H., van Loan C.~F., 1996. Matrix Computations, John Hopkins Uni. Press, Baltimore USA

\bibitem[Hurlburt et al.(1989)]{hpwb89}
Hurlburt N.~E., Proctor M.~R.~E., Weiss N.~O., Brownjohn D.~P., 1989, JFM,
207, 587

\bibitem[Kunz(2011)]{kunz11}
Kunz M.~W., 2011, MNRAS, 417, 602

\bibitem[Kunz et al.(2012)]{kbrs12}
Kunz M.~W., Bogdanovi\'{c} T., Reynolds C.~S., Stone J.~M., 2012, ApJ, in press

\bibitem[Lamb(1997)]{lamb97}
Lamb H., 1997, Hydrodynamics, Cambridge Uni. Press, Cambridge UK

\bibitem[McCourt et al.(2010)]{mpsq11}
McCourt M., Parrish I.~J., Sharma P., Quataert E., 2011, MNRAS, 413, 1295

\bibitem[Parrish et al.(2012)]{pmqs12}
Parrish I. J., McCourt M., Quataert E., Sharma P., 2012, 422, 704

\bibitem[Parrish \& Quataert(2008)]{pq08}
Parrish I.~J., Quataert E., 2008, ApJ, 677, L9

\bibitem[Parrish, Quataert \& Sharma(2009)]{pqs09}
Parrish I.~J., Quataert E., Sharma P., 2009, ApJ, 703, 96

\bibitem[Parrish, Quataert \& Sharma(2010)]{pqs10}
Parrish I.~J., Quataert E., Sharma P., 2010, ApJ, 712, L194

\bibitem[Peterson \& Fabian (2006)]{pf06}
Peterson J.~R., Fabian A.~C., 2006, Phys. Rep., 427, 1

\bibitem[Proctor \& Weiss(1982)]{pw82}
Proctor M.~E.~P., Weiss N.~O., 1982, RPPh, 45, 1317

\bibitem[Quataert(2008)]{quataert08}
Quataert E., 2008, ApJ, 673, 758

\bibitem[Ruszkowski \& Oh(2010)]{ro10}
Ruszkowski M., Oh S.~P., 2010, ApJ, 713, 1332

\bibitem[Rybicki \& Lightman(1979)]{rl79}
Rybicki G.~B., Lightman A.~P. 1979, Radiative Processes in Astrophysics. Wiley,
New York

\bibitem[Schekochihin et al.(2005)]{sckhs05}
Schekochihin A.~A., Cowley S.~C., Kulsrud R.~M., Hammett G.~W., Sharma P.,
2005, ApJ, 629, 139

\bibitem[Schekochihin et al.(2010)]{scrr10}
Schekochihin A.~A., Cowley S.~C., Rincon F., Rosin M.~S., 2010, MNRAS, 405, 291

\bibitem[Spitzer(1962)]{spitzer62}
Spitzer L.~Jr., 1962, Physics of Fully Ionized Gases. Wiley Interscience, New
York, NY

\bibitem[Stone et al.(2008)]{sgths08}
Stone J.~M., Gardiner T.~A., Teuben P., Hawley J.~F., Simon J.~B., 2008, ApJS,
178, 137

\end{thebibliography}
\end{document}